# CSCW PRINCIPLES TO SUPPORT CITIZEN SCIENCE

Julia Katherine Haines
Department of Informatics
University of California, Irvine
Irvine, CA, 92607  USA
hainesj@uci.edu

## ABSTRACT

Citizen science changes the way scientific research is pursued. It opens up data collection and analysis to the general public, to the wisdom of crowds. In this emerging area, there is much research to be done to better understand how we can develop citizen science infrastructure and continue the democratization of science. In creating such systems, there is much we can learn from principles that have emerged out of computer-supported cooperative work (CSCW) research. In this paper, I use a nine-step framework to highlight where CSCW knowledge can contribute.

## CSCW PRINCIPLES FOR CITIZEN SCIENCE

Citizen science has two main purposes: first, to aid in scientific research and secondly, to provide education to the public, as successful citizen science projects should ideally engage the community in science in addition to producing scientific results. In order to succeed in reaching these goals as well as the goals of scientific education and involvement, systems to support citizen science need to be evaluated in certain ways to continue improving design. In this paper, I explore how cyberinfrastructures for citizen science have been and can be evaluated and discuss some design recommendations and the methods by which these systems can be evaluated and refined.

### A Nine Step Framework

The Cornell Lab on Ornithology (CLO) has a model for how citizen science projects should be designed and implemented so that they can fulfill both research and education goals [3]. This 9-step model serves as a good framework to analyze both the benefits and challenges of collaborative science systems. For each step, I will present the most pertinent CSCW topics and discuss implications for design.

*1- Choose a scientific question*

In choosing a scientific question that can be addressed through cyberinfrastructure, it is important to think about how a particular scientific question can be translated and de- and re-contextualized in a way that the participant can understand. Though the research itself is highly complex, it must be able to be broken down into shared elements of understanding. Boundary objects work because they are understandable by both parties without lots of unnecessary or unhelpful additional information. [1] A second area of concern is the often widely-distributed geography and timespan of citizen science projects. Coupling is key. Only work that can be loosely coupled can be done successfully through citizen science crossing numerous locations. [5]

*2- Form a team*

Coupling is also important in forming a team of scientists, educators, technologists, and evaluators to work on the project. Because this requires a much higher mutual dependence for advancing the work, it should be more collocated. One important criterion is common ground. [5] For localized projects, common ground will likely be easily established. For projects that are more international in scope, rich media interfaces would help in building common ground and developing trust.

*3- Develop, test, and refine*

To develop a well-designed task protocol, we must be familiar with related organizations and industries. We must also understand the workflow at a finer granularity. Steiner's taxonomy of tasks [6] and demands is important to think about and consider. The task must be divisible, with a fairly low level of interdependence and high quantity of data, but a range of quality should be possible. Task design of any citizen science project should look not only at what tasks can be accomplished well through the law of large numbers, but also how those tasks can serve group well-being and provide support to contributors. Providing educational components and motivational elements is key, so that the project does not end up squandering the goodwill of the citizen scientists who end up participating.

*4- Recruit participants*

Finding participants and enticing them to continue participating is perhaps the most challenging step. Overcoming the challenge of critical mass [4] is essential to the success of citizen science. The bigger problem is perhaps sustained participation, not recruiting. One model to emulate is games with a purpose (GWAPs), which use play to accomplish another task. Whether a project is specifically designed as a game or not, incentives must be designed in such a way so as to encourage sustained contribution. In fact, the right kind of messaging about participants' contributions may be enough to create sustained motivation [2].

### 5- Train participants

Key criteria for collective wisdom are diversity, independence, decentralization, and aggregation. [7] Any form of training should take into account these needs and determine ways to best train participants to come out with a good aggregate product. Legitimate peripheral participation, which is "a theory of social learning" shows that novices in a community of practice may become experts over time through observation and gradual initiation. This framework presents training more as an opportunity to learn from each other rather than a simple step in the process of conducting a citizen science project.

### 6- Accept, edit, and display data

In accepting, editing, and displaying the data that is contributed, research on wikis might be most helpful. Due to the unverified nature of the data that is collected, a system of governance or a way to weed out useless or harmful data is needed. Whether this must be done in a centralized way or can in fact be done is a more decentralized way is dependent on the specific project, its volunteer base, and how long it has been going on. Designing with attribution in mind would allow credibility to be better evaluated and might create debate that is beneficial to the project. Identification could be automatic, but have an option to disable to preserve anonymity.

### 7- Analyze and interpret data

Because citizen science projects must rely on individual participants to take responsibility for creating the data, there are concerns about reliability in the analysis and interpretation of the output. The law of large numbers and the central limit theorem enable a level of statistical reliability. However, there are still concerns related to individual contributions. Ratings systems (as in [8]) would be one general recommendation for making sure the data analysis in citizen science system is solid.

### 8- Disseminate results

In terms of disseminating results to the participants involved, we must consider ways of making such dense and complex data understandable. Utilizing visualization techniques may help by showing patterns of activity and making the data easier to grasp. Visualization options may be particularly helpful in more complex citizen science projects where participant data analysis is encouraged.

### 9- Measure outcomes

Multiple methods of continued research would be helpful in understanding how to continue to develop citizen science infrastructure. More qualitative methods, particularly textual analysis of forums, wikis, and other data, would be useful in understanding the social structures and uncovering the way this collaborative work is actually done and it's educational outcomes. By conducting pre- and post-project surveys, looking at exchanges through features like wikis, and doing in-depth interviews, we can perhaps better understand all the benefits enabled through citizen science systems and uncover some of the problems that are being encountered.

## CONCLUSION

Beyond gaming formats, wikis, forums, and data visualizations, the citizen science projects of the future may benefit from tools and techniques not yet created. The possibilities for citizen science research are just beginning to be discovered; in order to determine the trajectory and how to evaluate and evolve these systems, we need to conduct more in-depth research into the social, technological, and work-process aspects of citizen science.